\documentclass{ametsocV6.1-archive}
\usepackage{graphicx}
\usepackage{hyperref}
\usepackage{floatrow}

\title{On the attribution of weather events to climate change using empirically-fit extreme value distributions}

\authors{Peter Sherman,\aff{a}\correspondingauthor{Peter Sherman, petersherman@g.harvard.edu}
Peter Huybers,\aff{a,b}
and Eli Tziperman,\aff{a,b} 
}

\affiliation{\aff{a}{School of Engineering and Applied Sciences, Harvard University, Cambridge, MA, USA}\\
\aff{b}{Department of Earth and Planetary Sciences, Harvard University, Cambridge, MA, USA}\\
}

\abstract{Changes in extreme weather events are a potentially important aspect of anthropogenic climate change (ACC), yet, are difficult to attribute to ACC because the record length is often similar to, or shorter than, extreme-event return periods. This study is motivated by the ``World Weather Attribution'' initiative (WWA) and, specifically, their approach of fitting extreme value distribution functions to local observations. They calculate the dependence of distribution parameters on global mean surface temperature (GMST) and use this dependence to attribute extreme events to ACC. Applying this method to preindustrial climate simulations with no time-varying greenhouse gas forcing, we still find a strong dependence of distribution parameters on GMST. This dependence results from internal climate variability (e.g., ENSO) affecting both extreme events and GMST. Therefore, dependence on GMST does not necessarily imply an effect of ACC on extremes. We further consider whether an extreme value, normal, or log-normal distribution better represents the data; if a GMST-dependence of distribution parameters is justified using a likelihood ratio test; and if a meaningful attribution is possible given uncertainties in GMST dependence. We find, for example, that an attribution of Australia's 2020--2021 Bushfires to ACC is difficult due to the effects of internal variability. For the 2019--2021 drought in Madagascar we find that the small number of available data points precludes a meaningful attribution analysis. Overall, we find that the effects of internal climate variability on GMST and the uncertain relationship between GMST and regional extremes may lead to inaccurate attribution conclusions using the part of the WWA approach examined here.}

\graphicspath{{Figures/}{./}}
\begin{document}

\maketitle

\section{Introduction}

The Intergovernmental Panel on Climate Change \cite[][]{IPCC-2021:WG-I-chapter3, IPCC-2023:synthesis-section2, IPCC-2021:WG-I-SPM} found that global mean surface temperature (GMST) between 2011--2020 was 1.09 ${}^\circ$C [0.95 ${}^\circ$C--1.20 ${}^\circ$C] above preindustrial (1850--1900) levels. Although climate change is clearly manifested in GMST warming, and there are reasons to expect extreme events to change with mean climate, it is a complex task to attribute individual extreme weather events to anthropogenic climate change (ACC). Extreme events---i.e., heatwaves, droughts, and heavy precipitation---are rare by definition, so the observational record may be too short to detect anthropogenic signals. This issue is especially pervasive in lower-income countries where data are sparse, and the historical record is short \cite[][]{Otto-Harrington-Schmitt-et-al-2020:challenges}. The situation is further problematic for extreme precipitation event attribution, which is highly sensitive to the specific timescales and percentiles used to define extreme precipitation events \cite[][]{Pendergrass-2018:what}.

The field of attribution of extreme events has progressed significantly over the past decade and encompasses several approaches \cite[see example reviews and cited references within,][]{Stott-Gillett-Hegerl-et-al-2010:detection, Hegerl-Zwiers-2011:use, Sciences-2016:attribution, Clarke-Otto-Stuart-Smith-et-al-2022:extreme, Philip-Kew-Van-et-al-2020:protocol}. The approaches that have been applied include the fraction of attributable risk \cite{Allen-2003:liability, Stott-Stone-Allen-2004:human}, using simulations from climate models, using constraints derived from optimal fingerprinting techniques \cite[][]{Christidis-Stott-Zwiers-2015:fast}, using climate models to validate observed probability density functions \cite[PDFs,][]{Perkins-Pitman-Holbrook-et-al-2007:evaluation, Lobeto-Menendez-Losada-2021:future}, and the ``Analogue-Based Approach'' \cite[][]{Lorenz-1969:atmospheric, Yiou-Vautard-Naveau-et-al-2007:inconsistency, Vautard-Yiou-2009:control, Faranda-Bourdin-Ginesta-et-al-2022:climate}. Another methodology fits statistical distributions to the historical record (again, either observed or from climate models) in what \cite{Stott-Christidis-Otto-et-al-2016:attribution} refer to as ``Empirical-Based Approaches.'' This approach assesses how the return period of particular events has changed over the historical record, with the purpose of detecting potential trends in the frequency of extreme events. These statistical approaches have been used to study trends in extreme events such as flooding in Thailand in 2011 \cite[][]{Van-Van-Allen-2012:absence}. Approaches using climate models allow for a better estimate of the magnitude and direction of human-induced contributions to an extreme event than is possible using observations alone.

The climate model approaches detailed in the above paragraph can also be performed with Single Model Initial-condition Large Ensembles (SMILEs), composed of a climate model run several times with slightly different initial conditions. Averaging over multiple climate model ensemble members can reduce the signal from modes of internal variability, which are typically uncorrelated between ensemble members, facilitating the identification of forced climate change signals. This averaging generally improves the detection of trends in the magnitude or frequency of an extreme event, which are difficult to quantify in a single simulation or in the observational record where the signal from internal variability may be dominant \cite[][]{Yoshida-Sugi-Mizuta-et-al-2017:future, Yamaguchi-Chan-Moon-et-al-2020:global}. These simulations have been used in the context of tropical cyclone slowdown \cite[][]{Yamaguchi-Chan-Moon-et-al-2020:global}, as well as drought events in South and East Africa \cite[][]{Lott-Christidis-Stott-2013:can, Pascale-Kapnick-Delworth-et-al-2020:increasing}. The attribution literature is extensive, and for further detailed discussion, we refer the reader to the above reviews and to recent studies \cite[][]{Stott-Christidis-Otto-et-al-2016:attribution, Knutson-Kossin-Mears-et-al-2017:detection, Seneviratne-Zhang-Adnan-et-al-2021:weather}.

One particular approach that is the focus of this paper and that has been used in recent years by WWA as a rapid response to extreme events attempts to fit the observed data to an extreme value PDF \cite[][]{WWA-webpage}. The fit allows for a dependence of the distribution parameters on explanatory variables, typically the GMST \cite[e.g.,][]{Ciavarella-Cotterill-Stott-et-al-2021:prolonged, Oldenborgh-Krikken-Lewis-et-al-2021:attribution, Harrington-Wolski-Pinto-et-al-2022:limited, Philip-Kew-Van-et-al-2020:protocol}. A GMST dependence is then interpreted as a response of extreme events to ACC. This dependence is then used to attribute extreme events by quantifying the effects of anthropogenic forcing on the probability of their occurrence. Technically, this is achieved by writing the location parameter of the extreme event distribution, $\mu$, as a function of $\overline{T}$, the (low-pass filtered, via a 4-year running average) GMST ``as a measure of anthropogenic climate change'' \cite[][]{Harrington-Wolski-Pinto-et-al-2022:limited}, such that $\mu=\mu_0+\alpha\overline{T}$. The GMST-dependence, $\alpha$, is estimated via a fit to observations. The fitted PDF is then used to evaluate the probability of extreme events at 1900, $P(\mu_0+\alpha\overline{T}(1900))$ relative to those at 2020, $P(\mu_0+\alpha\overline{T}(2020))$. This allows for estimating how the probability of an extreme event changed between these times. A change in the probability of an extreme event is inferred by the WWA approach if $\alpha$ is non-zero, based on the change in GMST. 

Although WWA does not carry out formal hypothesis testing, we interpret that they adopt a null hypothesis of $\alpha$ equal to zero implying no ACC influence on extremes and an alternative hypothesis that, if $\alpha$ is not equal to zero, ACC does influence extremes. This issue is discussed in detail in section~\ref{sec:methods}\ref{sec:null-hypotheis} and further addressed throughout the text.

Our focus is on the empirically-based approaches used by WWA, but it should be noted that WWA also requires evidence from climate model simulations before making any attribution statements. WWA often uses multiple climate models and is careful to exclude models based on their performance and return periods relative to observations \cite[e.g.,][]{Pinto-Barimalala-Philip-et-al-2023:extreme}.  In this study we use a single model (CESM) to demonstrate some of our points but focus on assessment of observational methodologies associated with fitting of extreme value distributions to observations.  Our goal is limited to evaluating the observational approach described by WWA and suggesting additional tests to examine its robustness, as opposed to developing alternative attribution approaches.

Motivated by these pioneering efforts based on the fit to extreme event PDFs, we reexamine three test cases analyzed by WWA: the Siberian heat wave analysis of \citet[][hereafter CC2021]{Ciavarella-Cotterill-Stott-et-al-2021:prolonged}, the analysis of high temperatures associated with the Australian Bushfire by \citet[][hereafter OK2021]{Oldenborgh-Krikken-Lewis-et-al-2021:attribution}, and the Madagascar drought analysis of \citet[][hereafter HW2022]{Harrington-Wolski-Pinto-et-al-2022:limited}. These cases were chosen as they represent a range of applications, including the attribution of heatwaves at high and low latitudes (Siberia vs.\ Australia) and of an extreme drought. Each of these studies leverages either a Generalized Extreme Value (GEV) or Generalized Pareto Distribution (GPD) fit to assess potential contributions to extreme events from ACC. WWA studies use both climate model simulations and observational analysis, and we focus on the observational component. In each of the three cases we use the CESM Large Ensemble \cite[][]{Kay-Deser-Phillips-et-al-2015:community} to put the observed instrumental record into perspective, and we explore potential caveats concerning the methods examined and suggest how these limitations might be addressed.

The issue highlighted by our analysis is simple: internal variability modes such as ENSO affect both local temperature extremes and GMST. In some cases, the effect on local extremes (say the effects of ENSO on heatwaves in Australia) may be large, but the effect on GMST is small. This might lead to a false conclusion that a small GMST signal leads to a comparatively large effect on local extremes. To demonstrate this point, we analyze a preindustrial climate model run with no anthropogenic forcing. We find, for example, that the fit of Australian heat extremes leads to values of $\alpha$ that are substantially greater than zero in the absence of ACC because of the effects of internal climate variability modes such as ENSO on both GMST and on local extremes. Our findings show that $\alpha$ can be strongly affected by internal variability, biasing the attribution results and complicating the WWA approach. 

Our study addresses four questions: (1) Is a GMST dependence of the PDF parameters necessarily an indication of the signal of ACC? (2) Is a GEV or GPD necessary to fit the available observations, or are more standard normal  or log-normal distributions sufficient? (3) Is the addition of GMST-dependent distribution parameters, which form the basis for the attribution analysis, justified statistically, and is the uncertainty range of these parameters, specifically $\alpha$, sufficiently small to allow for attribution? (4) What do SMILEs tell us about the amount of data needed to determine the GMST dependence with confidence? WWA uses multiple attribution tools, including analysis of simulations from multiple models and observations \cite[][]{Philip-Kew-Van-et-al-2020:protocol}. We focus only on the latter; specifically, on the fitting of extreme value distributions to observations. Our goals are limited to evaluating this observational approach and suggesting additional tests to examine its robustness, and do not include developing alternative attribution approaches.

The rest of the paper is structured with Section~\ref{sec:methods} describing the statistical models, observation datasets, and methodologies. Section~\ref{sec:results}a shows that a GMST-dependence of distribution parameters does not necessarily represent the effects of ACC using an analysis of a CESM preindustrial simulation (i.e., without time-varying anthropogenic and natural forcings). We then further examine the three attribution cases in the remainder of section \ref{sec:results}. Section~\ref{sec:conclusions} provides a summary and discussion of the results.

\section{Methods and Data}
\label{sec:methods}

In this section, we briefly describe the extreme value distributions used in the following sections (\ref{sec:methods-GEV} and \ref{sec:methods-GPD}) and discuss uncertainty estimates and empirical cumulative distribution functions (CDFs) that we will use later (\ref{sec:methods-uncertainty}). We describe the deviance statistic test which we use to examine the justification for adding GMST-dependent parameters (\ref{sec:methods-deviance}), and the mean residual life plot used to test the justification for using a GPD and its threshold value (\ref{sec:MRL}). The data used are described in section (\ref{sec:methods-data}).

\subsection{The Generalized Extreme Value (GEV) distribution} 
\label{sec:methods-GEV}

Extreme value distributions are a family of PDFs that can be used to represent the statistical behavior of block maxima or minima of a record. An example is an annual time series of the maximum daily temperature at each year, the ``block'' being one year in this case. It can be shown that the distribution of such maxima follows one of three classes of PDFs: Gumbel, Fr\'echet, or Weibull \cite[][]{Coles-Bawa-Trenner-et-al-2001:introduction}, depending on the shape of the tail of the distribution of the events for which block maxima are calculated. Rather than explicitly specifying one of these three classes in fitting data for extreme events, one can combine these distributions into one functional form, represented by the Generalized Extreme Value (GEV) distribution. Large quantities of data are typically needed to accurately fit a GEV and calculate its parameters \cite[e.g.,][]{Philip-Kew-Van-et-al-2020:protocol, Trevino-McKinnon-Huybers-2020:extremely}.

The PDF of a variable $x$ whose statistics are governed by the GEV is given by
\begin{equation}
    GEV(x,\mu,\sigma,\xi) = \frac{1}{\sigma}t(x,\mu,\sigma,\xi)^{\xi+1}e^{-t(x,\mu,\sigma,\xi)},
    \label{eq:GEV1}
\end{equation}
where 
\begin{equation}
    t(x,\mu,\sigma,\xi) = {\left(1+\xi\frac{x-\mu}{\sigma}\right)}^{-1/\xi},
    \label{eq:GEV2}
\end{equation}
under the assumption that $\xi \neq 0$. If $\xi = 0$, $t(x,\mu,\sigma,\xi)$ is defined as,
\begin{equation}
    t(x,\mu,\sigma,\xi) = e^{-\frac{x-\mu}{\sigma}}.
    \label{eq:GEV3}
\end{equation}

In some of the attribution studies we follow for this paper (specifically, CC2021 and OK2021), the location parameter $\mu$ is assumed to vary linearly with the 4-year smoothed GMST $\overline{T}$, as,
\begin{equation}
    \mu = \mu_0 + \alpha \overline{T}.
    \label{eq:GEV-mean-and-GMST}
\end{equation}
Estimation of the parameter $\alpha$ is the key to the attribution of extreme events. However, this determination may be limited due to the short observational record \cite[][]{Zwiers-Zhang-Feng-2011:anthropogenic, De-Giugni-Pugliese-et-al-2018:gev}. To estimate the mean $\mu_0$, scale parameter $\sigma$, the GMST dependence $\alpha$, and the shape parameter $\xi$, we follow \cite{Coles-Bawa-Trenner-et-al-2001:introduction} and WWA \cite[][]{Philip-Kew-Van-et-al-2020:protocol} and maximize the log-likelihood of the data points $x_t$ using a GEV distribution,
\begin{align}
    l(\mu_0,\sigma,\xi,\alpha) 
    &= \sum_{t=1}^m \log\left(GEV(x_t,\mu_0,\sigma,\xi,\alpha)\right).
\end{align}
We apply this approach, following CC2021 and OK2021, to observations and to data from the CESM Large Ensemble \cite[][]{Kay-Deser-Phillips-et-al-2015:community}.

\subsection{Generalized Pareto Distribution (GPD)}
\label{sec:methods-GPD}

In some applications, one is interested only in points in a time series that exceed a specified threshold, whose statistics are represented by the Generalized Pareto Distribution \cite[][]{Coles-Bawa-Trenner-et-al-2001:introduction}. In order to study the extreme drought events in Madagascar using a record of precipitation, HW2022 fit a Generalized Pareto Distribution (GPD) to the bottom 20th percentile of precipitation, i.e., years associated with drought. This low tail was converted to a high tail by multiplying the precipitation data by $-1$. Although this converts the lowest precipitation values to maxima, the transformed data have a hard maximum at zero. We note that the validity of using a GPD, in this case, is questionable because no such hard maximum is allowed by---or represented in the functional form of---the GPD. 

The PDF of the GPD is given by 
\begin{equation}
    GPD(x,u,\sigma,\xi) = \frac{1}{\sigma}\left(1+\xi\frac{x-u}{\sigma}\right)^{-(\frac{1}{\xi}+1)}.
\end{equation}
HW2022 assumed that the threshold parameter, which they denote $u$ \cite[denoted $\mu$ by][]{Philip-Kew-Van-et-al-2020:protocol} and scale parameter $\sigma$ both depend exponentially on the 4-year running-averaged GMST, $\overline{T}$ such that,
\begin{align}
    u &= u_0\exp({\alpha \overline{T}}/{u_0}), \nonumber\\
    \sigma &= \sigma_0\exp({\alpha \overline{T}}/{u_0}).
   \label{eq:GPD-exponential-mu-sigma}
\end{align} 
The exponential fit was assumed by HW2022 in order to ensure that the scaling factor $\exp(\alpha \overline{T}/u_0)$ applied to both $u_0$ and $\sigma_0$ is strictly positive and maintain a fixed ratio between the mean and variance. We obtain the optimized GPD parameters ($u_0,\sigma,\xi,\alpha$) by again maximizing the log-likelihood
\cite[Eq. 4.10,][]{Coles-Bawa-Trenner-et-al-2001:introduction},
\begin{equation}
    l(u,\sigma,\xi) = \sum_{t=1}^m \log(GPD(x_t,u,\sigma,\xi)),
\end{equation}
using the trust-region constrained algorithm in the Scipy Python package (`trust-constr' method).

The GPD requires the bound $\sigma > 0$ and consistency relations of the form $x\geq u$ (for $\xi\geq 0$), and $u\leq x$ and $x\leq u-\sigma/\xi$ (for $\xi<0$). As the optimization iteratively searches for the optimal parameter values, these constraints are occasionally violated, leading to complex numbers or NaNs (not a number) in the likelihood. The GPD optimization is notably more prone to such failures and sensitive to the initial guesses of $\alpha,u_0,\xi,\sigma$ than the GEV optimization. In order to avoid such search failures, we replace complex and NaN terms in the log-likelihood sum with large penalty terms whenever they occur due to constraints being violated. In addition, if the optimization fails, we re-run the optimization again with new randomly selected initial guesses until convergence to a solution that maximizes log-likelihood and satisfies the constraints is found. We note that related solutions have been used to deal with GEV/GPD convergence issues by \cite{Robin-Ribes-2020:nonstationary}.

\subsection{Uncertainty estimates and empirical cumulative distribution functions}
\label{sec:methods-uncertainty}

We follow WWA studies and use non-parametric bootstrapping to estimate uncertainty ranges for the optimized parameters. In particular, we use the uncertainty in $\alpha$ as one way to evaluate the validity of the attribution results. We also compare the appropriately fitted distributions (GEV and normal distributions for extreme temperatures or GPD and log-normal for droughts) to the empirical cumulative distribution functions (CDFs) of the observations to assess the quality of the fit. The empirical CDF is calculated by sorting extreme values (either temperature or precipitation) from smallest to largest, calculating the cumulative sum, and normalizing to a maximum CDF value of one. We calculate error bars in the empirical CDFs based on bootstrap resampling with replacement of the observations. The empirical CDF is calculated for each sampling, and we then calculate the 5--95th and 2.5--97.5th percentile ranges from the resulting distribution at each value of the variable whose CDF is estimated. In order to estimate uncertainty in the parameters estimated from our GEV or GPD maximum likelihood fit, we similarly use bootstrap resampling. We sample with replacement the GMST and corresponding extreme temperature record 5,000 times and optimize the parameters for each resampling. We then estimate the 90th and 95th percentiles from 5,000 bootstrap samples. For the Madagascar precipitation case, we perform 5,000 bootstrap resamples when analyzing the observations, but 1,000 bootstrap resamples when analyzing the CESM Large Ensemble due to the slow convergence and expensive computations using the GPD log-likelihood maximization.

\subsection{Deviance Statistics (Likelihood Ratio Test)}
\label{sec:methods-deviance}

In general, when fitting a model to data, if the number of fit parameters is increased, one expects a better fit because of the greater flexibility of the fitted model. Specifically, in the case considered here, adding a parameter $\alpha$ to include the GMST dependence of the location/threshold parameter will increase the maximum log-likelihood of the data. The salient question is whether the improvement in the fit exceeds that expected simply on account of the larger number of model parameters. To evaluate the improvement of the fit, \citet[][theorem 2.7]{Coles-Bawa-Trenner-et-al-2001:introduction} recommend the use of a deviance statistic, also known as a likelihood ratio test. This test uses a measure, $D$, equal to twice the difference of the sum of log-likelihoods of the model with $k$ additional parameters minus that of the simpler model.  Because more parameters lead to a better fit and thus higher log-likelihood, $D$ is non-negative.  $D$ is then compared to a $\chi_k^2$ distribution to assess if the $k$ additional parameters significantly improve the likelihood. For example, with $k=1$, the 95th percentile of $\chi_1^2$ is equal to 3.8.  If $D$ is larger than this threshold, we conclude that adding an additional parameter significantly improves the fit. It can then be inferred that the added parameter---in our specific case, the GMST dependence, $\alpha$---meaningfully explains features of the data. On the other hand, if the deviance statistic is insignificant, one concludes that the addition of another parameter to the statistical fit is not justified by the data and, therefore, that this parameter cannot be used to draw conclusions about the data. Specifically, if one finds that the addition of $\alpha$ is not justified, the implication is that the estimated value should not be used to calculate the effects of climate change on the extreme events under examination. We also estimate the distribution and uncertainty of $D$ through bootstrap resampling of the data 5,000 times.

\subsection{Mean residual life plot}
\label{sec:MRL}

In the process of fitting a GPD, a minimum threshold is chosen to select the data that represent extreme values that should be fit. The GPD fit can be sensitive to the selection of this threshold, where too low of a threshold could violate the asymptotic nature of the GPD, and too high of a threshold would provide only a few relevant data points and lead to an unstable fit. In order to determine an appropriate threshold, one leverages what is known as the mean residual life (MRL) plot. The MRL uses the GPD-based expectation value of the data
\cite[Eq. 4.8,][]{Coles-Bawa-Trenner-et-al-2001:introduction}.
For a given threshold, $u_0$, the expectation value (mean) of the data $X$ that exceed the threshold can be written for $\xi < 1$ as,
\begin{equation*}
    E(X-u_0|X>u_0) = \frac{\sigma_{u_0}}{1-\xi}.
\end{equation*}
It can be shown that for all thresholds $u > u_0$, this expectation value is 
\cite[Eq. 4.9,][]{Coles-Bawa-Trenner-et-al-2001:introduction},
\begin{equation}
    E(X-u|X>u) = \frac{\sigma_{u_0}+u\xi}{1-\xi}.
    \label{eq:MRL}
\end{equation}
That is, for $u > u_0$, $E(X-u|X>u)$ is a linear function of $u$. This expected linearity of Equation~\ref{eq:MRL} with $u$---for values of $u$ for which the GPD fit is appropriate---can be used to test the validity of the chosen threshold used in the GPD analysis via a plot which we use in our results section known as the MRL plot.

\subsection{Hypothesis testing}
\label{sec:null-hypotheis}

Although WWA does not carry out formal hypothesis testing, we interpret that they adopt a null hypothesis of $\alpha$ equal to zero implying no ACC influence on extremes and an alternative hypothesis that, if $\alpha$ is not equal to zero, ACC does influence extremes.  This inference is based on the quote, ``If the one- or two-sided 95\% confidence interval [for the change in probability, which has the same effect as $\alpha$] excludes zero, the change is statistically significant'' \cite[see Section 4.3.1 of][]{Philip-Kew-Van-et-al-2020:protocol}. We assess whether rejection of a null hypothesis of $\alpha$ equal to zero permits accepting an alternate of ACC influencing extreme events. In other words, we assess if a nonzero $\alpha$ necessarily means a role for ACC. We note that another null hypothesis for purposes of heat could be that the location of PDF of heat extremes is shifted by an amount equaling the GMST warming, as would be represented by $\alpha=1$.  For purposes of consistency with the null implied by WWA, however, we only consider a null of $\alpha$ equal to zero.

WWA uses two variants of their analysis approach, one in which the extreme event in question is incorporated into the estimate of the PDFs, which we follow here, and one in which it is excluded \cite[][]{Philip-Kew-Van-et-al-2020:protocol}. The act of choosing to study a class of events immediately following the occurrence of a particularly extreme example is known to induce a bias related to whether it is included or excluded \cite[][]{MIRALLES2023100584}. We choose to include the extreme value in question, which tends to make the parametric dependence on GMST larger and tends to make it easier to reject the null of no change in favor of the alternate hypothesis.  Within the context of the WWA hypothesis test, this choice to include the extreme in question makes it more likely to conclude that AAC significantly influences extreme events.

\subsection{Data}
\label{sec:methods-data}

We briefly review the observations and our processing used in each of the three case studies reconsidered here. In addition, we describe the model Large Ensemble and preindustrial simulations that we analyze.

\paragraph{Siberian Heatwave.} Following CC2021, we rely on two sets of observations: (1) daily maximum temperature extremes for each year from station data in Verkhoyansk as discussed in CC2021; and (2) 4-year smoothed GMST anomalies from GISTEMP \cite[][]{Lenssen-Schmidt-Hansen-et-al-2019:improvements}. CC2021 also performed a GEV analysis on temperature anomalies over a region in Siberia, which we do not focus on here in order to study specifically the extreme values detected at the Verkhoyansk station. At the time of the CC2021 study, the daily maximum temperature extreme in 2020 at the Verkhoyansk station was recorded as 38$^\circ$C. This temperature extreme currently appears as a missing data point on the National Climatic Data Center website (\href{https://www.ncei.noaa.gov/cdo-web/search}{https://www.ncei.noaa.gov/cdo-web/search}, station RSM00024266, VERHOJANSK, data downloaded Aug 2023) but we still use this unconfirmed value in our analysis. 

We also study GMST anomalies and extreme temperatures over the Siberian region defined in CC2021 within the CESM Large Ensemble \cite[][]{Kay-Deser-Phillips-et-al-2015:community} over the period 1926--2019. The CESM Large Ensemble contains 40 simulations of the climate over the period 1920--2100 under historical and RCP 8.5 external forcing, of which we use the 35 that were run on the NCAR supercomputer. These simulations differ only in small perturbations to the initial conditions, which lead to different temporal sequences of modes of internal variability. We use model output only up to the year 2020, and, as a result, there is little sensitivity to the emissions scenario used \cite[DAMIP,][uses RCP 4.5]{Gillett-Shiogama-Funke-et-al-2016:detection} because emissions scenarios only significantly diverge later. We do not use some additional runs done on the University of Toronto supercomputer that contain a global mean temperature bias \cite[][]{CESM_LENS_Issues}.

In addition, we study GMST anomalies and extreme temperatures within a 1,800-year CESM preindustrial simulation, yielding 18 segments of 94 years, chosen for consistency with the length of the observed record.   This preindustrial simulation uses fixed greenhouse gas concentrations, volcanic activity, anthropogenic aerosols, land use, and solar forcing all set to their respective levels in the year 1850.  Most importantly in the context of this study is the lack of time-varying anthropogenic forcing in this simulation. 

\paragraph{Australian Bushfire Heatwave.} Following OK2021, we again use two sets of observations for the years 1920--2019: (1) annual (July--June) maxima of 7-day moving average daily-maximum surface temperature data from the Australian Water Availability Project (AWAP) over the Southeastern region in Australia defined in OK2021; and (2) 4-year smoothed GMST anomalies from GISTEMP. We apply a similar analysis to the CESM Large Ensemble and to the long CESM preindustrial run as for the Siberian Heatwave case except, in this case, using 18 segments of 98 years for consistency with the observations.

\paragraph{Madagascar Drought.} Following HW2022, we use two datasets. The first data are 2-year means of precipitation from ERA5 \cite[][]{Hersbach-Bell-Berrisford-et-al-2020:era5} over the region of Madagascar that was defined in HW2022. HW2022 explained that they used a ``24-month running mean rainfall data from July to June.'' Our approach is to calculate total annual precipitation and then calculate a running 2-year average of this annual data resulting in a new smoothed annual time series. For 1951--2020 the bottom 20\% precipitation 2-year periods are selected for the analysis, corresponding to the following 14 two-year periods: 1956--1957, 1957--1958, 1958--1959, 1959--1960, 1962--1963, 1990--1991, 1991--1992, 1992--1993, 2008--2009, 2009--2010, 2015--2016, 2016--2017, 2017--2018, 2019--2020. The second data are 4-year smoothed GMST anomalies from GISTEMP. We note that the analysis following HW2022 relies on a fit to only 14 data points. We discuss the issues that are involved in fitting extreme distribution functions to such a small number of data points below. We also study the equivalent variables within the CESM Large Ensemble over the period 1950--2019, consistent with the period of observations. We again apply a similar analysis to the CESM Large Ensemble and to the long CESM preindustrial run, in this case using 18 segments of 70 years.

We note that in each of these three case studies, WWA leverages many datasets, including ERA5 reanalysis, station observations, and the multimodel CMIP5 ensemble. We use a subset of the data that WWA study but also the CESM Large Ensemble data in a consistent manner across the three example cases that we evaluate.

\section{Results}
\label{sec:results}

We first describe our analyses of how extreme events depend on GMST in the context of the long CESM pre-industrial simulation in subsection~\ref{sec:results}\ref{sec:control-run-analysis}. The simulation contains no time-varying anthropogenic forcing, making it useful for evaluating a null-hypotheses associated with no anthropogenic influence. We then consider each of the three case studies in subsections \ref{sec:results}\ref{sec:results-Siberia}, \ref{sec:results}\ref{sec:results-Australia}, and \ref{sec:results}\ref{sec:results-Madagascar}. For each case we assess how GEV or GPD distributions fit the observations relative to normal or log-normal distributions. We then evaluate the justification for the inclusion of a GMST dependence of the distribution parameters using the deviance statistic and estimate the uncertainty level of the parameter $\alpha$ used for the attribution. Finally, we examine the amount of data needed to adequately constrain GMST dependence in a GEV or GPD fit.

\subsection{Extremes analysis of a preindustrial simulation}
\label{sec:control-run-analysis}

WWA examines the dependence of the distribution parameters of extreme events on the 4-year running mean GMST. Commonly a linear dependence of a location parameter on GMST, as represented by the parameter $\alpha$, is used to estimate the PDF of extreme events in, for example, 1900 versus 2020.  Changes in the PDF between these epochs are interpreted as owing to the effects of anthropogenic climate change, implying that $\alpha$ indicates the extent to which anthropogenic climate change affects extremes.  We show here that it is possible, however, for internal climate variability, such as ENSO, to produce covariance between GMST and extremes that is expressed as a non-zero $\alpha$ even without anthropogenic influence. This covariance is demonstrated most clearly in this section by using a pre-industrial climate model run, but we also elaborate upon this basic result using the CESM Large Ensemble. For attribution purposes this results suggests that the value of $\alpha$ calculated in the presence of a GHG increase can be affected by internal variability, biasing the attribution results. It should be noted that in a realistic scenario $\alpha$ is affected by both anthropogenic climate change and internal variability. The scenario of internal variability alone analyzed here is artificial, of course, and does not capture that, in the current climate, GMST is a reasonable proxy for anthropogenic climate change. Analyzing simulations that only include internal variability nevertheless provides insight into the effects of internal variability in more complete scenarios.

\begin{figure}[!tbp]
\centering
\includegraphics[width=0.75\textwidth]{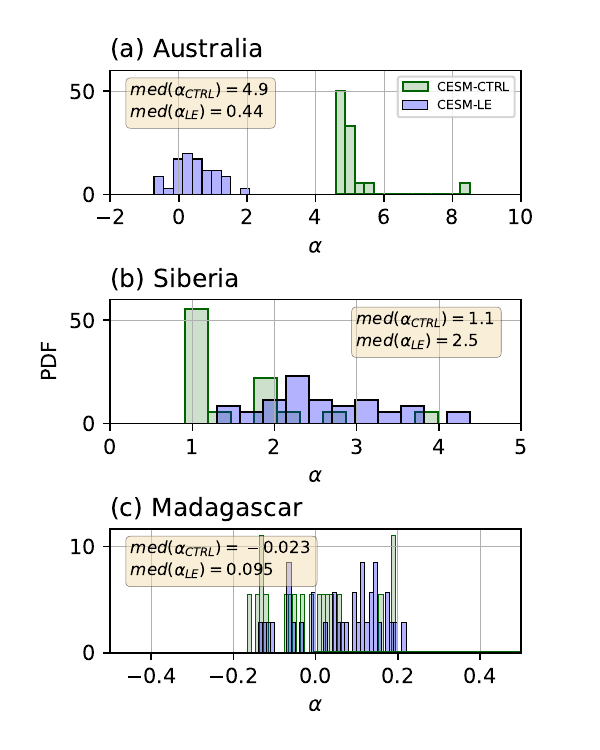}
\caption{\textbf{Effect of internal climate variability on} $\alpha$ \textbf{distribution.} PDFs of $\alpha$ for (a) The Australian 7-day moving mean of the maximum temperature related to the analysis of the 2019--2020 bushfires. (b) The Siberian maximum daily temperatures that were used in the analysis of the 2020 heatwave. (c) The Madagascar 2-year precipitation used in the analysis of the drought of 2019--2021. The analyses are based on the 1,800-year CESM preindustrial run with 94-, 98-, and 70-year segments, respectively (green), and the 35 Large Ensemble members (blue). The lengths of these segments correspond to the extent of the observed records used in CC2021, OK2021, and HW2022. The median values for the CESM preindustrial and Large Ensemble for each of the cases are indicated in the boxes. The median values are Australia: $\alpha_{CTRL}=4.9$, $\alpha_{LE}=0.4$; Siberia: $\alpha_{CTRL}=1.1$, $\alpha_{LE}=2.5$; Madagascar: $\alpha_{CTRL}=-0.02$, $\alpha_{LE}=0.09$.}
\label{fig:alpha_distributions_CTRL}
\end{figure}

To examine this potential role of internal variability, we start with the case of the hot weather associated with the bushfires in Southeastern Australia and perform a GEV analysis using a 1,800-year CESM preindustrial run. This run holds anthropogenic CO$_2$, other anthropogenic greenhouse emissions, volcanic (natural) emissions, land use changes, and anthropogenic aerosols at values estimated for the year 1850.  We divide the run into 18 data segments of length 94 years, equal to the length of the observed record and analogous to having 18 simulations of historical climate without ACC.  The distribution of $\alpha$ values for the GEV preindustrial analysis is shown by the green bars in Fig.~\ref{fig:alpha_distributions_CTRL}a. We find a large and positive $\alpha$ calculated from essentially all model output segments centered around a median value of 4.9 $^\circ$C per $^\circ$C, or that extreme summer temperatures increase by 4.9 $^\circ$C for every 1 $^\circ$C GMST warming.  Note that, following WWA, we are using four-year-running average GMST values for this computation.  

It is difficult to imagine a physical causal mechanism that would lead to such an enormous amplification of the GMST signal in Australian heatwaves. A more plausible interpretation is that internal variability --- for example, ENSO --- could increase both extreme temperatures over Australia and, to a smaller degree, GMST, generating covariance between extreme events and GMST. In this scenario GMST is not an indicator of forcing of local extremes; rather, both local extremes and GMST are influenced by a third factor. It is known that several modes of internal variability affect extreme temperatures in Australia, including ENSO, the Indian Ocean Dipole, and the Southern Annular Mode \cite[][]{Hendon-Thompson-Wheeler-2007:Australian, Ummenhofer-England-McIntosh-et-al-2009:what, King-Pitman-Henley-et-al-2020:role}. \citet{Min-Cai-Whetton-2013:influence} find that ENSO and the Indian Ocean Dipole can lead to warming of extreme temperatures in Southeastern Australia by around $1^\circ$C, indicating that internally generated variability can have an effect that is similar in magnitude to ACC.

WWA recognizes that $\alpha$ may be influenced by internal variability and, as noted, use a 4-year running mean GMST to filter out the effects of short-term internal variability.  This approach, however, may not fully remove the ENSO signal.  Furthermore, there is also the potential for nonlinear interactions between ENSO and GMST as well as the presence of slower modes of variability such as the Pacific Decadal Oscillation (PDO). 

A similar analysis of Australian heat extremes made using the CESM Large Ensemble, which includes anthropogenic forcing, gives a distribution of $\alpha$ that is closer to zero (blue bars in Fig.~\ref{fig:alpha_distributions_CTRL}) than in the unforced simulations (green bars). This indicates that both ACC and internal variability influence regional extremes and GMST, such that estimates of $\alpha$ will reflect multiple processes. In simulations with time-varying anthropogenic forcing, however, there is no way to disentangle the effect of internal variability on $\alpha$ using the WWA approach as described in published literature. We conclude that the magnitude of $\alpha$ may be a biased indicator of anthropogenic influence. Including indices of internal variability as covariates (by adding parameters that represent the dependence of the distribution on regional climate variability modes such as ENSO) may help to disentangle their respective contributions. Yet the projection of internal variability onto GMST can still confound the results, and adding additional parameters to the distribution to represent the dependence of the distribution on regional climate variability modes such as ENSO may not be justified given that below we find that even adding the single parameter $\alpha$ can be questioned due to the short observed record.

These results highlight a potential major difficulty with analyses based on a fit of distribution parameters to GMST in a more realistic scenario that includes both internal variability and ACC: the calculated $\alpha$, and therefore the shift in the probability of extreme events, may be due to internal variability rather than only ACC. When analyzing the observational record in section \ref{sec:results-Australia}, we find $\alpha\approx2\;^\circ$C per $^\circ$C. This suggests a two-degree warming of extreme temperatures for a one-degree warming of the GMST, which is at the upper end of the range of values estimated from CESM LE. Based on our results from the CESM preindustrial run with a time-invariant anthropogenic forcing, we therefore suspect that the estimate obtained for $\alpha$ from the observed record is biased high by internal variability.

Analysis of the CESM preindustrial run for the case of Siberian heatwaves (green bars in Fig.~\ref{fig:alpha_distributions_CTRL}b) shows that the value for $\alpha$ deduced from a record similar in length to that of the observations ranges between 1 to 4, although there is again no time-varying anthropogenic forcing in the corresponding model run. This large and variable response implies that a non-zero $\alpha$ does not necessarily imply an effect of ACC. It may suggest that natural variability affects the value of $\alpha$ for Siberian heat waves during some decades but not others. This analysis shows that $\alpha$ equal to zero is an unsuitable null hypothesis for the purposes of evaluating the presence of an anthropogenic influence on extreme events (for context, we note again that WWA does not explicitly define a null hypothesis).

For the case of Madagascar Drought (Fig.~\ref{fig:alpha_distributions_CTRL}c) we find a large spread in $\alpha$ for both the preindustrial run as well as the CESM Large Ensemble, with both centered near zero. The fact that the $\alpha$ distribution from the preindustrial run is centered around zero suggests that a null hypothesis of $\alpha$ equal to zero may be appropriate, but the fact that $\alpha$ from the Large Ensemble is also centered around zero indicates that the there is little potential for this test to discern an ACC influence on extremes.  That is, low extremes of precipitation show no systematic increase or decrease related to GMST in either the forced or unforced runs such that this test has low statistical power.  We further consider this case below using observations.

The analysis in this paper shows that internal variability can lead to a large $\alpha$. We conclude that in a more typical scenario analyzed using the WWA methodology and applied to observations under varying anthropogenic GHG and aerosols, the value of $\alpha$ may still be affected by internal variability and, therefore, may bias the estimates of the effect of ACC on the return time of extreme events. One could justify the WWA use of the 4-year averaged GMST as being a proxy for known greenhouse gas concentrations. If CO$_2$ were indeed used, the analysis of the preindustrial run would, of course, show no dependence, eliminating this problem, though then introducing issues associated with lags between GHG concentrations and warming \cite[][]{Proistosescu-Huybers-2017:slow}. Because WWA studies use GMST, we follow that methodology here.

The result that internal variability may strongly affect $\alpha$ when the 4-year average GMST is used to link extreme events to ACC suggests that it might be useful to reevaluate some of the WWA results based on the fit to extreme distributions. It is possible, of course, that a strong anthropogenic signal in a few decades could significantly affect extreme events and overwhelm the effects of internal variability in cases such as the Siberian Heatwave. The attribution of extreme temperatures may be more difficult in cases like the Australian example, where the effects of internal variability are large, as indicated by the large $\alpha$ detected in the CESM preindustrial simulation. In such cases, it would be important to correctly identify all relevant covariates --- such as the possible effect of ENSO in the Australian case --- and note that misleading results for $\alpha$ may be obtained if one relies only on GMST.

\subsection{Siberian Heatwave}
\label{sec:results-Siberia}

We now explore other facets of the attribution of the Siberian Heatwave in the summer of 2020 to anthropogenic change, notwithstanding the foregoing finding that the interpretation of $\alpha$ is complicated by the presence of internal variability. We focus specifically on the CC2021 analysis of the June daily maximum temperatures because it allows us to address the question of whether the data justify the use of non-stationary (GMST-dependent) extreme value distributions.

While it is understood based on theoretical considerations that a GEV is the appropriate representation of block-maxima of data \cite[][]{Coles-Bawa-Trenner-et-al-2001:introduction}, it is not obvious for a given problem what minimum block size justifies using a GEV, and specifically if a year of daily maximum data suffices. This is especially an issue given that the temperature is auto-correlated and, therefore, more data are needed per averaging block to justify the use of a GEV. It is, therefore, worthwhile to check how well other standard distributions perform in fitting the data.

\begin{figure}[tp]
\centering
\includegraphics[width=1\textwidth]{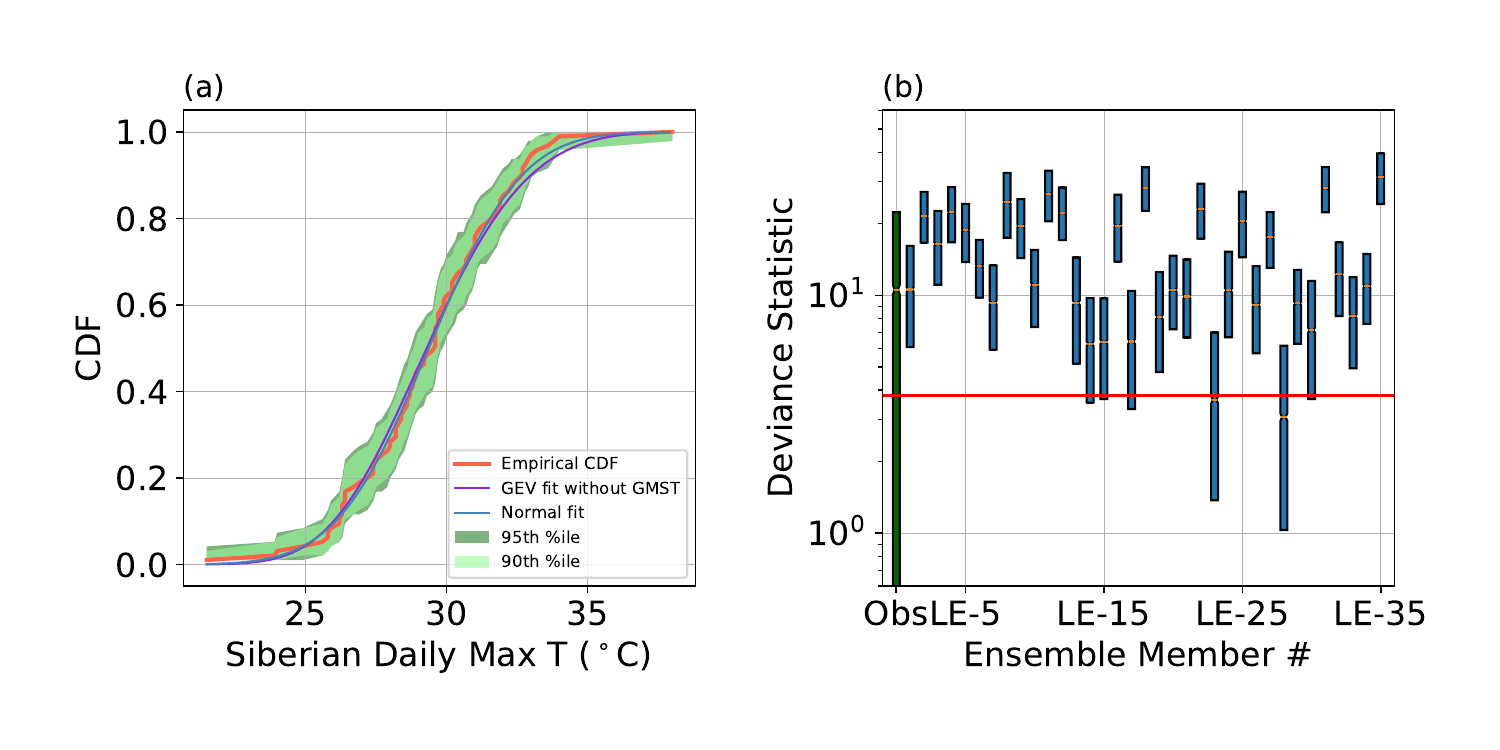}
\caption{\textbf{Siberian Heatwave: (a) Examining the fit of a GEV and normal distribution to Verkhoyansk station data.} The red curve indicates the empirical CDF of the highest daily maximum temperature for each year for the Verkhoyansk station data. The purple and blue lines denote fits by a GEV (with a constant location parameter) and by a normal distribution, respectively. The light and dark green shadings correspond to 90th and 95th percentile confidence intervals (section~\ref{sec:methods}\ref{sec:methods-uncertainty}). \textbf{(b) Justification for using a GMST-dependent mean.} Box plots of the deviance statistic between a GEV distribution without a GMST-dependent mean relative to a GEV distribution with a GMST-dependent mean using the Siberian heatwave data with 5,000 bootstrap resamples. The edges of the boxes correspond to the first and third quartiles. The deviance statistic for the station observations is shown in green as the leftmost box. The deviance statistic was also computed for each of the 35 members from the CESM Large Ensemble (blue). The 95th percentile significance level from a $\chi_{k=1}^2$ distribution with one degree of freedom is shown by the horizontal red line, corresponding to the addition of one degree of freedom by including $\alpha$ in the fitted model.}
\label{fig:Siberia_GEV_fit_and_deviance}
\end{figure}

A cumulative distribution estimate of the highest June daily-maximum temperature observed each year between 1926--2020 at the Verkhoyansk station is shown in Fig.~\ref{fig:Siberia_GEV_fit_and_deviance}a along with a 95\% confidence interval of the CDF obtained by bootstrapping. In the first step of our analysis, we consider cumulative distribution function fits to the observed distribution, assuming the data follows either a GEV or normal distribution. The fitted distributions are specified here to be stationary (by setting $\alpha=0$ and $\mu=\mu_0$ in Eqns.~\ref{eq:GEV1}--\ref{eq:GEV-mean-and-GMST}). Fits are obtained by maximizing the likelihood of the data, and the results are shown in Fig.~\ref{fig:Siberia_GEV_fit_and_deviance}a. Both distributions generally fit the observations well, with the normal fit completely within the 95\% range of the observed distribution, and with the GEV fit being good except at the highest temperature, where it slightly underestimates the probability of obtaining the highest temperatures. There is only one data point above 35 \textdegree{}C, so the empirical CDF is poorly constrained there in any case. Interestingly, the normal distribution and GEV CDFs look similar and lie within the confidence intervals defined for the empirical CDF, despite the normal distribution having fewer parameters than the GEV. It is worth noting that a GEV and a normal distribution may assign very different probabilities to extreme events, even when the two distributions provide an adequate fit to the observed CDF. These modeling decisions are therefore of critical importance for attribution studies, which usually focus as much or more on estimation of the exceedance probability and return period of an event --- and, consequently, on the changing likelihood and fraction of attributable risk of an event --- as they do on the change in intensity. It is therefore useful to consider alternate distributional forms.

The quality of fits indicates that an extreme value distribution is not empirically a better choice than a normal distribution for fitting the data.  Similarly, a non-stationary distribution with $\alpha\ne0$ may not be necessary given that the stationary distributions seem to provide a satisfactory fit to the empirical cumulative distributions, and we further examine this next. CDFs are commonly used to differentiate between distribution functions, and the results of a Kolmogorov-Smirnov test (Supplementary Fig.~SI-1 for normal, GEV and empirical distributions) also suggests that the normal distribution is consistent with the observed data.   An application of Occam's razor principle suggests that the simpler normal distribution should be used rather than a GEV if there is not a significant improvement in the distributional fit with the GEV. This is also discussed by \cite{Philip-Kew-Van-et-al-2020:protocol} who mention explicitly that the extreme value distributions may not outperform a simpler (normal) distribution in some cases. \cite{Otto-Philip-Kew-et-al-2018:attributing} opted to use a normal distribution in lieu of a GPD for attribution analysis of a cold spell in Peru because the normal ``uses all data, rather than the GPD, which samples only the cold tail''. CC2021 found that a normal distribution fits the data well but proceeded to use a GEV for the bulk of the analysis. These results question the need for extreme value distribution functions when it comes to attribution analysis. One might want to use proper selection criteria (the Akaike Information Criterion, AIC, or the deviance statistic AKA the log likelihood ratio test which we use in this work, or similar) to rigorously decide which of the models performs best.

\begin{figure}[!tbhp]
\begin{center}
\includegraphics[width=0.8\textwidth]{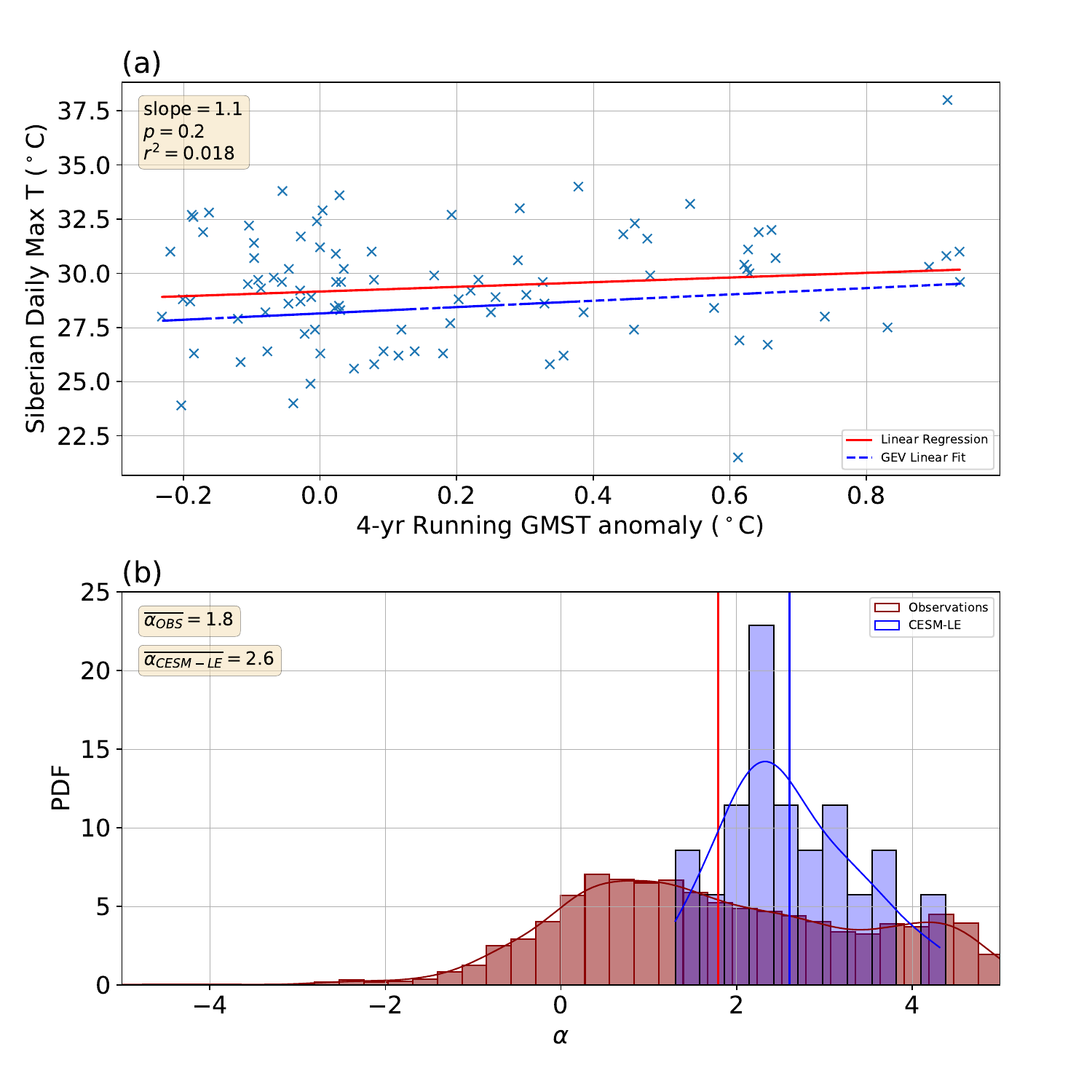}\\
\end{center}
\caption{\textbf{Siberian Heatwave: uncertainty in $\alpha$.} (a) Scatter plot of the daily max temperature at the Verkhoyansk station against GMST. The red line indicates a linear regression fit to the data, and the blue line indicates a linear fit using the location parameter and $\alpha$ derived from the GEV maximum likelihood estimate. (b; red) PDF of the GEV-fit $\alpha$ distributions from the station data, using 5,000 bootstrap resamplings of the observed time series. The  $\alpha$ value from the observations (no resampling) is shown by the vertical red line. (b; blue) Shown is a histogram of the GEV-fit $\alpha$ PDFs derived from the CESM Large Ensemble, where the $\alpha$ values are calculated from each ensemble member with no resampling. The average $\alpha$ value across the Large Ensemble is indicated with the blue line. There are 94 annual data points for each ensemble member.}
\label{fig:siberia_alpha_distributions}
\end{figure}

We now proceed to examine the role of GMST in the fit to a GEV distribution (eq.~\ref{eq:GEV-mean-and-GMST}) in this particular application, following CC2021. A linear regression of GMST anomalies against daily max temperatures gives a slope of 1.1 $^\circ$C per $^\circ$C (Fig.~\ref{fig:siberia_alpha_distributions}a), where this value indicates that the annual maximum daily temperatures at Verkhoyansk increase slightly more quickly than GMST. The squared Pearson correlation, however, is only $r^2=0.01$, with a p-value of 0.2, indicating that GMST is only a weakly significant indicator of trends in Siberian temperature maxima.

Fig.~\ref{fig:siberia_alpha_distributions}b shows the maximum likelihood fit of a nonstationary GEV, including $\alpha$, for multiple bootstrapping samples as specified in the Methods section. The values of $\alpha$ estimated from the observations is 1.8 $^\circ$C per $^\circ$C, larger than the slope of 1.1 obtained from the linear regression. This discrepancy between $\alpha$ and the slope suggests that the conclusions from an attribution analysis can be highly sensitive to the choice of the fitted distribution, again underscoring the importance of model selection. The [2.5\%, 97.5\%] confidence interval is [$-1.2$, 4.8], while the 5\% value is $-0.7$ (Fig.~\ref{fig:siberia_alpha_distributions}b; red plots). Both indicate that $\alpha$ is not significantly different from zero. It is noteworthy that the red distribution in Fig.~\ref{fig:siberia_alpha_distributions}b is bimodal, presumably because some of the bootstrapped samples included the 2020 event and some did not. This suggests that the fitted distribution is very sensitive to the inclusion or exclusion of the event (again consistent with \cite{MIRALLES2023100584}); that is, $\alpha$ would be significantly non-zero if the event is included, but is not significant when it is omitted. While CC2021 did not provide an uncertainty range for $\alpha$, they note that the change in intensity of Siberian heatwaves at a fixed probability is estimated from their inferred PDF as 1.04 $^\circ$C (with a 95\% confidence interval from 0.35 to 3.4). $\alpha$ is directly related to the change in intensity: the change in intensity is $k * \alpha$, where $k$ is the observed GMST change between the two periods of interest, usually 1.2$^\circ$C. Assuming a GMST change of 1.2$^\circ$C, we infer the value of $\alpha$ in CC2021 is 1.3 with a 95\% confidence interval from 0.42 to 4.1. The central estimate of $\alpha$ is similar to our results but entails a smaller uncertainty range, likely due to our decision to include the 38$^\circ$C event in our bootstrap resampling, but which was not done in CC2021. The bootstrapped distribution for all GEV parameters for the observations is given in Supplementary Fig.~SI-2.

The deviance statistics, $D$, (section~\ref{sec:methods}\ref{sec:methods-deviance}) indicate that the addition of a model parameter, $\alpha$, representing the GMST dependence, is justified in the case of the Verkhoyansk observations,  ((Fig.~\ref{fig:Siberia_GEV_fit_and_deviance}b, leftmost bar is above the threshold indicated by the red horizontal line). Yet the deviance statistic uncertainty range includes a large range that is below the significance line, which seems appropriate given that the uncertainty range for $\alpha$ includes negative values.

To provide context for the interpretation of the Verkhoyansk 
observational results, we also examine output from the CESM Large Ensemble, which includes time-varying anthropogenic forcing, at the grid point containing the Verkhoyansk station. We find that all ensemble members within the CESM Large Ensemble indicate a positive value of $\alpha$, with a mean value of 2.60 $^\circ$C per $^\circ$C (Fig.~\ref{fig:siberia_alpha_distributions}b; blue plots). Similarly, 33 of 35 ensemble members give a deviance statistic that supports the use of a non-stationary component (that is, a GMST-dependence) in the GEV statistic (Fig.~\ref{fig:Siberia_GEV_fit_and_deviance}b).

\begin{figure}[t]
\centering
\includegraphics[width=0.7\textwidth]{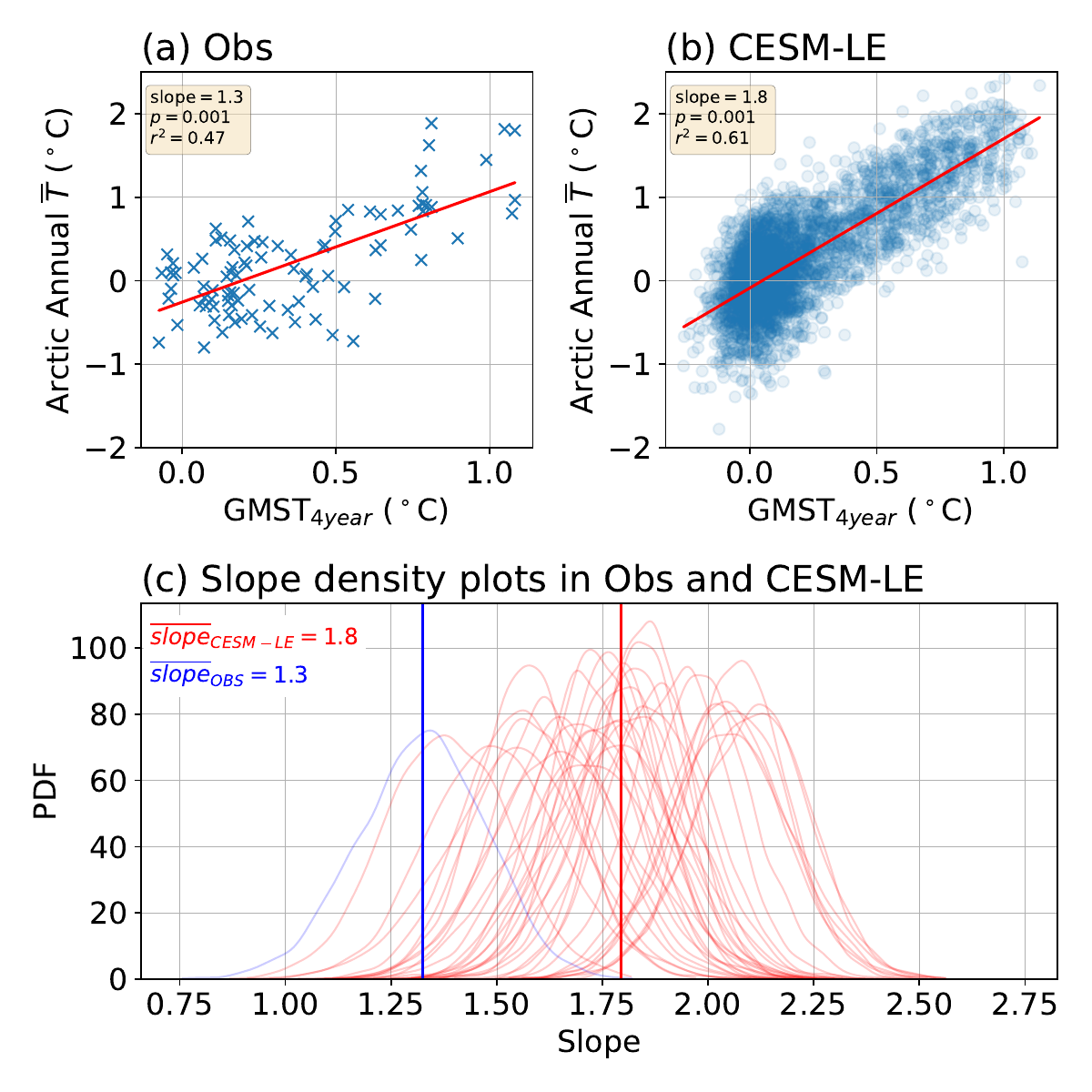}
\caption{\textbf{Internal variability and the attribution of Siberian heatwaves.} (a) A scatter plot of the annual average JJA Arctic (north of $70^\circ$ N) surface temperatures against the 4-year running mean of GISTEMP annual GMST. (b) Same, for the 35 members of the CESM Large Ensemble over the period 1926--2019. (c) PDF of the slopes of annual average JJA Arctic surface temperatures against the 4-year running mean of annual GMST within each member of the CESM Large Ensemble with 5,000 bootstrap resamples (red). The same distribution is indicated for GISTEMP observations in blue. The vertical red line indicates the mean slope across the CESM Large Ensemble, and the vertical blue line indicates the slope determined from observations.}
\label{fig:GMST_ArcticT}
\end{figure}

The fact that the LE results indicate a stronger link between Siberian extremes and GMST (that is, a larger $\alpha$) is consistent with the LE, on average, simulating an Arctic-wide summer warming rate of 1.8 $^\circ$C per $^\circ$C warming in GMST, whereas an observational analysis only indicates a rate of 1.3 $^\circ$C Arctic JJA warming per $^\circ$C in GMST (using GISTEMP for both; Fig.~\ref{fig:GMST_ArcticT}a,b). This greater amplification of Arctic warming may be related to the fact that CESM1 has a relatively high climate sensitivity of 4.1 K \cite[][]{gettelman2019high}. That said, one expects heatwaves to be strongly influenced by the JJA climatology, and Fig.~\ref{fig:GMST_ArcticT}c shows that different ensemble members, corresponding to different realizations of internal variability, show very different dependencies of the Arctic JJA climatology on GMST. It is difficult, therefore, to estimate from the single realization corresponding to the observed record to what degree the Arctic JJA climatology is affected by ACC versus internal variability, adding to the uncertainty in attributing Siberian heatwaves. In order to address this issue, the WWA protocol includes a step where the same analysis is applied to climate models, leveraging a range of climate models so that results are not dependent entirely on one GCM, and results are reported from both sources.

\begin{figure}[t]
\centering
\includegraphics[width=0.9\textwidth]
{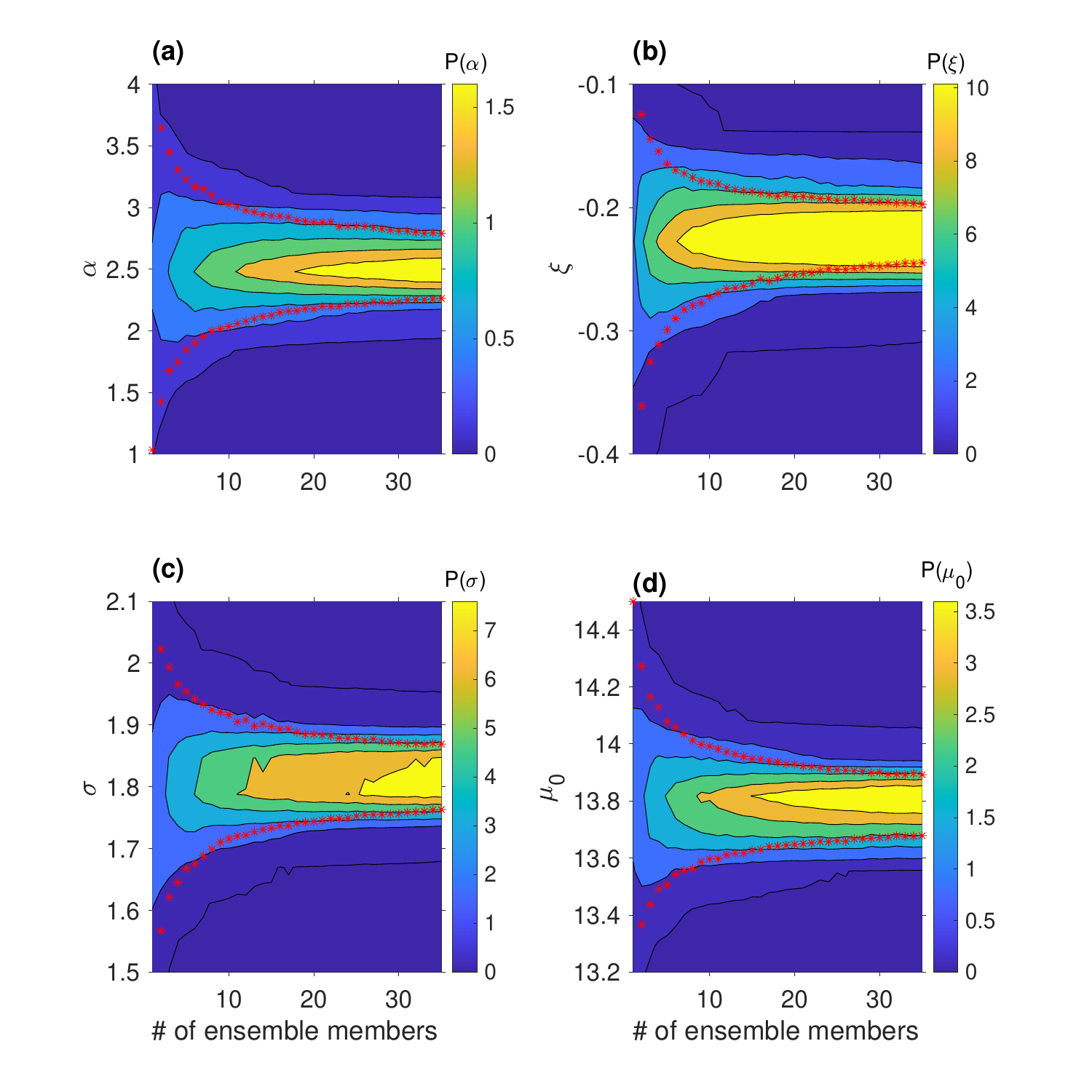}
\caption{\textbf{Siberian Heatwave: uncertainty in GEV parameters as a function of the number of ensemble members used.} Probability density function contour plots for each of the GEV parameters as a function of the number of ensemble members used from the CESM Large Ensemble and the parameter value. One ensemble member is equivalent in length to the historical record. Panel (a) represents the PDF for $\alpha$, (b) $\xi$, (c) $\sigma$ and (d) $\mu_0$. The PDFs were derived from the highest daily maximum Siberian temperature in June, as in CC2021, using the CESM Large Ensemble. Contours are drawn every 0.3 for (a), 2 for (b), 1.5 for (c), and 0.7 for (d), all starting at 0.05. We denote the 95\% confidence range for each number of ensemble members by the red asterisks.}
\label{fig:siberia_temp_parameters}
\end{figure}

Unsurprisingly, if data from multiple ensemble members are analyzed simultaneously as a single dataset, which mimics the availability of more data, the distribution of the GEV parameters narrows, generally converging around the mean value of the parameters across individual ensemble member fits (Fig.~\ref{fig:siberia_temp_parameters}). The 95\% confidence range for $\alpha$ (red asterisks in panel a) decreases from a large range of [1,4] ($^\circ$C per $^\circ$C) with one ensemble member to a smaller [2,3] range with 10 ensemble members. Similar behavior is seen for the other GEV parameters. This implies that with a sufficiently long record, one can reduce the uncertainty in the GEV parameters, as expected, but that might require much more data than is expected to be available in the historical record anytime soon. 

The overall result for the GMST dependence for the Siberian Heatwave analysis, which is the basis of the attribution to ACC, is largely consistent with the results found from CC2021 but with three important caveats.  First, we found in the previous subsection that internal variability can lead to GMST dependence that biases the value of $\alpha$ and hence the attribution results (Fig.~\ref{fig:alpha_distributions_CTRL}a).  Second, whereas a GEV fit a priori make sense to fit to block maxima of daily temperature extremes, a normal distribution with $\alpha = 0$ does not appear to be ruled out by the observations (Fig.~\ref{fig:Siberia_GEV_fit_and_deviance}a). This may be a result of both the relatively short observational record and the autocorrelated nature of the temperature extremes. Still, a non-zero $\alpha$ is found to meaningfully improve the statistical fit to the CESM Large Ensemble Siberian extremes based on deviance statistics.  Finally, the uncertainty range for $\alpha$, which was not explicitly evaluated in the original WWA analysis, includes negative values (Fig.~\ref{fig:siberia_alpha_distributions}b).

\subsection{Australian Bushfire Heatwave}
\label{sec:results-Australia}

\begin{figure}[t]
\centering
\includegraphics[height=8cm]{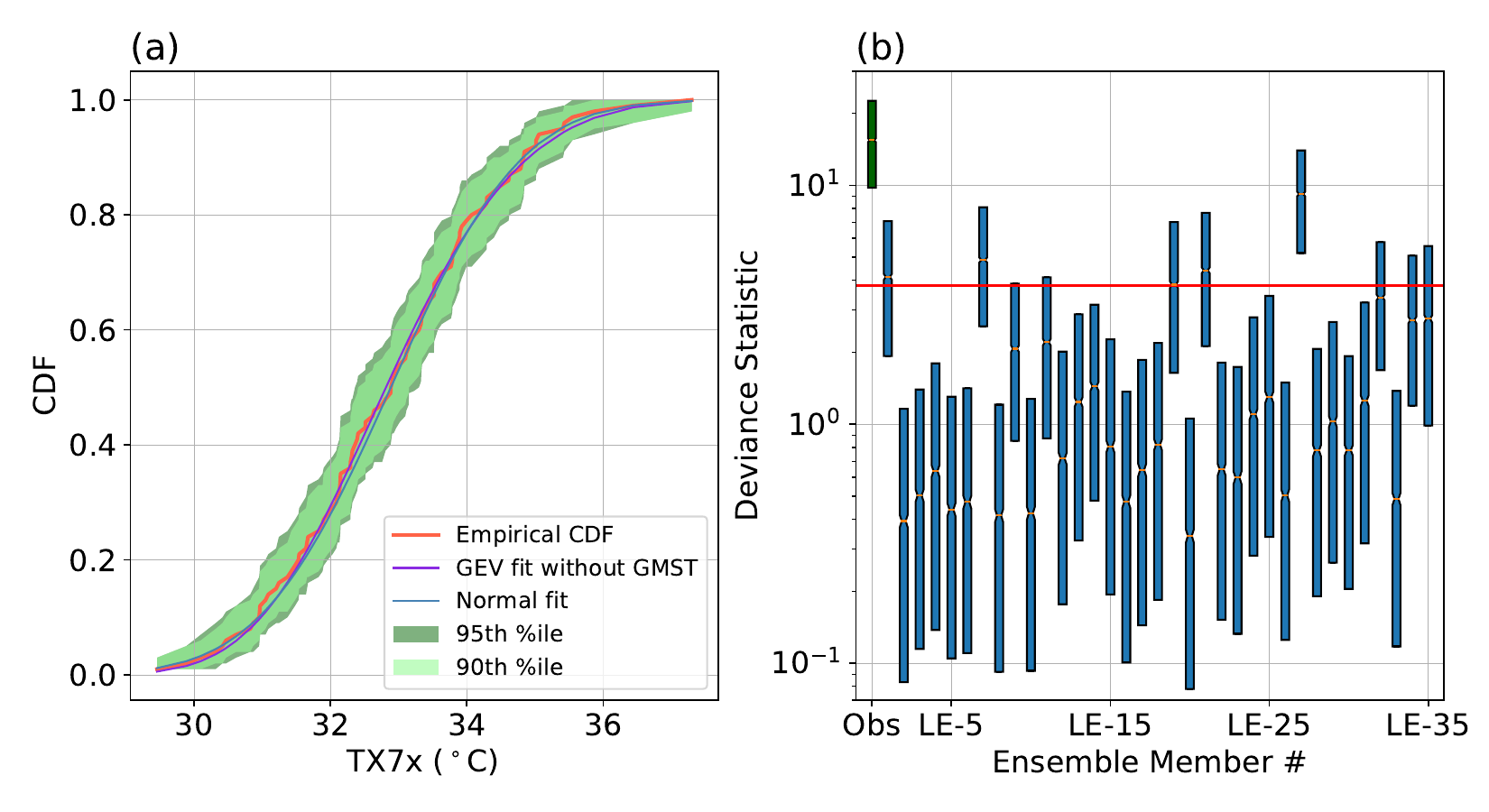}
\caption{\textbf{Australian Bushfire Heatwave: (a) Testing the justification for a GEV.} The red curve indicates the empirical CDF of the highest annual 7-day running mean of daily maximum temperature over the area defined in \cite{Oldenborgh-Krikken-Lewis-et-al-2021:attribution} for observations. The purple and blue lines denote GEV (with a constant mean), and normal distribution fits, respectively. The light and dark green shadings correspond to 90th and 95th percentile confidence intervals. \textbf{(b) Justification for using a GMST-dependent mean.} Box plots of the deviance statistic between a GEV without a GMST component and a GEV with a GMST component using the Australia Bushfire heatwave example with 5,000 bootstrap resamples. The edges of the boxes correspond to the first and third quartiles. The deviance statistic was computed for each of the 35 members from the CESM Large Ensemble (blue) and compared with the 95th percentile significance level from a $\chi^2$ distribution with one degree of freedom (red line). A box plot for the deviance statistic for observations is shown in green.}
\label{fig:australia-CDF-and-deviance}
\end{figure}

We now consider the analysis of possible connections between the hot weather that may have contributed to the Australian Bushfires of the summer of 2020--2021 and ACC, using both observations and the CESM Large Ensemble, following the observational analysis of OK2021. We again find that a normal distribution and a GEV distribution (both with a constant mean/location parameter) fit the 7-day moving mean of daily temperature maxima in the observations without a GMST dependence well (Fig.~\ref{fig:australia-CDF-and-deviance}a). In this case, a normal distribution fit to the empirical distribution has a mean absolute error (MAE) of 0.01, equal to the MAE for the GEV fit. Supplementary Fig.~SI-3 shows results for normal, GEV, and empirical distributions that again suggest that the normal distribution aligns well with the observed data.  The alignment may reflect that the small number of available data do not constrain the PDF very well. Given that the normal distribution has one less parameter, it seems to offer a simpler approach to representing the data. The deviance statistic for the observations does suggest a significant GMST component (leftmost bar in Fig.~\ref{fig:australia-CDF-and-deviance}b). However, given the possibility that this component is driven by internal variability rather than by anthropogenic change (section~\ref{sec:results}\ref{sec:control-run-analysis}, Fig.~\ref{fig:alpha_distributions_CTRL}a), one must conclude again that even if $\alpha$ significantly differs from zero, this might reflect a bias introduced by the effects of internal variability.  This underscores the need to include covariates representing relevant internal variability based on our physical understanding. The deviance statistic test for the CESM Large Ensemble shows that in only 5 of the 35 ensemble members is the extra complexity of adding GMST dependence to the distribution justified (Fig.~\ref{fig:australia-CDF-and-deviance}b). This raises the question of whether the influence of GMST, and therefore of ACC in the WWA formulation, is a necessary part of the statistical description of extreme temperatures in Australia as represented in the CESM Large Ensemble. It should be noted that our results relate only to one climate model that, of course, does not necessarily represent all the relevant physical processes at work in the real world. The use of multiple climate models in the WWA analysis helps ensure a wider range of process representations are included. 

\begin{figure}[!tbhp]
\centering
\includegraphics[width=0.8\textwidth]{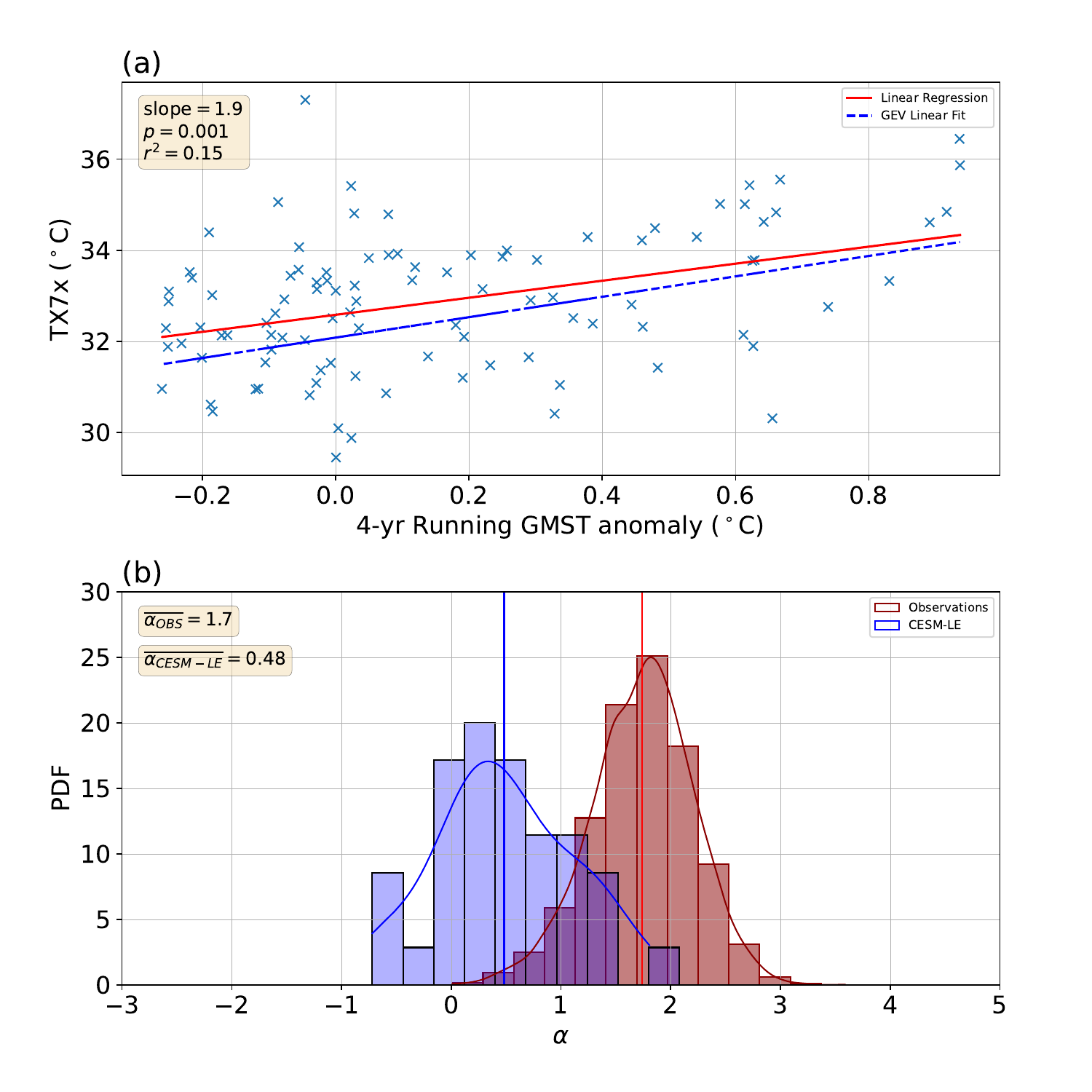}
\caption{\textbf{Australian Bushfire Heatwave: distribution of $\alpha$ values.} (a) Scatter plot of the highest annual 7-day running mean of daily maximum temperatures against GMST \cite[following][]{Oldenborgh-Krikken-Lewis-et-al-2021:attribution}. The red line indicates a linear regression fit to the data, and the blue line indicates a linear fit using the location parameter and $\alpha$ derived from the GEV maximum likelihood estimate. (b; red) PDF of the GEV-fit $\alpha$ distributions from observations, using bootstrap resampling. The $\alpha$ value from the observations without resampling is shown by the red line (at 1.7) (b; blue) Effect of the internal variability and number of data from a model large ensemble on the uncertainty in $\alpha$ for the Australian Bushfire heatwave example. Shown is a histogram of the GEV-fit $\alpha$ PDFs derived from the $\alpha$ value from each ensemble member with no resampling. The mean $\alpha$ value across the Large Ensemble is indicated with the blue line.}
\label{fig:australia_alpha_distributions}
\end{figure}

The regression analysis of Australian extreme temperatures and GMST, (Fig.~\ref{fig:australia_alpha_distributions}a) as well as the range of $\alpha$ values estimated from observations (Fig.~\ref{fig:australia_alpha_distributions}b; red) both show a GMST dependence near 2 $^\circ$C change in extreme temperatures per degree increase in GMST. As mentioned above, a sensible null hypothesis for heatwaves in a warming climate is that the extreme values shift with the mean warming \citep{Tziperman-2022:global}. The GMST dependence calculated here suggests that Australian heatwaves increase at nearly twice the rate of the GMST. Such a strong response requires a physical mechanism. Some candidates are (1) the fact that land heats up more quickly than the ocean, so GMST always tends to lag behind land surface temperatures; (2) soil-drying can lead to further increases in warming over land as less water is available for evaporative cooling, etc. It is not clear that these can lead to a doubling of the GMST effect but further exploration of this amplification is beyond the scope. The alternative is that this is a result of internal variability --- possibly, ENSO --- affecting both Australian heat extremes and the GMST, consistent with Fig.~\ref{fig:alpha_distributions_CTRL}a. Again, this suggests that using this value of $\alpha$ for attribution may lead to biased results. Finally, Fig.~\ref{fig:australia_alpha_distributions}b (blue) shows a large range of GMST-dependencies for different ensemble members in the CESM Large Ensemble, again highlighting that a large component of this dependence most likely represents internal variability rather than anthropogenic change. The distribution of all GEV parameters for the observations is given in Supplementary Fig.~SI-4. 
How the GEV parameters scale with the number of CESM Large Ensemble members used is indicated in Supplementary Fig.~SI-5.

\subsection{Madagascar Drought}
\label{sec:results-Madagascar}

Finally, we consider the attribution analysis for the 2019--2021 drought in Madagascar, following the lead of HW2022. This case differs from the two examined above in that it analyzes precipitation extremes that are positive by definition. HW2022, therefore, used a different extreme value distribution function, the GPD, which is meant to represent the high tail of the distribution of extreme events over a specified threshold \cite[][]{Coles-Bawa-Trenner-et-al-2001:introduction}. As discussed in Section~\ref{sec:methods}, HW2022 analyzed the driest 20\% of the 2-year averages of precipitation data from 1951--2020 multiplied by minus one to turn minima (droughts) into maxima. We follow their example, with the important caveat made in Section~\ref{sec:methods}\ref{sec:methods-GPD} that the validity of using a GPD this way for precipitation data is not guaranteed, given that there is a hard maximum at zero precipitation that seems inconsistent with the definition of the GPD. We also note that taking the lowest 20\% of 2-year precipitation data points amounts to only 14 data points, likely too low a number for a reliable attribution.

\begin{figure}[tp]
\centering
\includegraphics[width=0.8\textwidth]{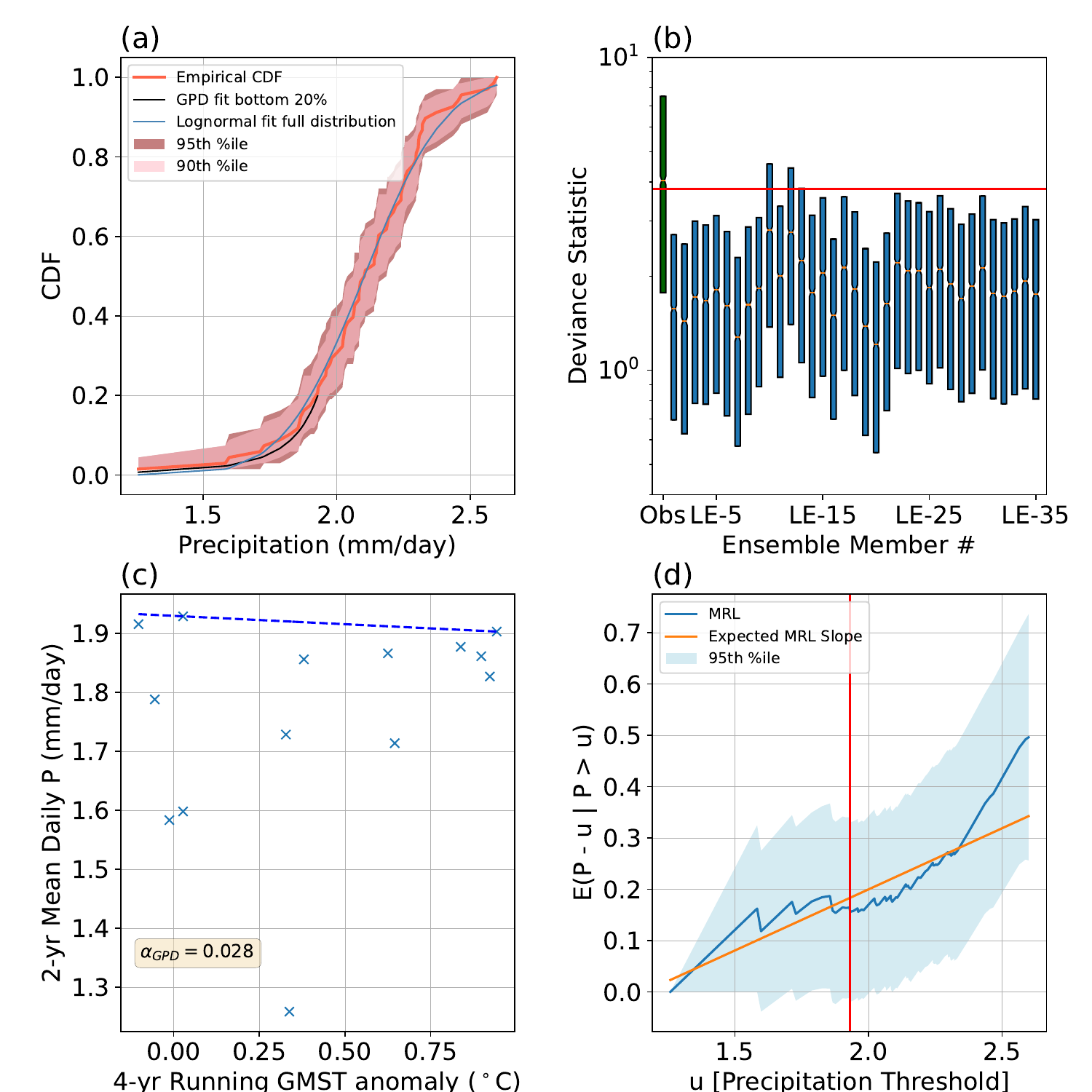}
\caption{\textbf{Madagascar Drought:} (a) Evaluating GPD and log-normal fits to rainfall data. An empirical CDF of rainfall is shown (red line, with 90th and 95th percentile confidence intervals shown by red shadings), based on 24-month running means of daily precipitation over the Madagascar area defined in HW2022 using ERA5 precipitation. A GPD fit to the bottom 20th percentile precipitation (black) and a log-normal fit to the full distribution (blue). \textbf(b) The deviance statistic between a GPD with and without a GMST component in the CESM Large Ensemble (blue) and observations (green). (c) Scatter plot of the 2-year running mean of ERA5 daily precipitation over the Madagascar area defined in HW2022 against the 4-year running mean of annual GMST for the lowest 20 percent precipitation data. The dashed blue line shows the exponential GPD log-likelihood fit for the threshold parameter for the observations without resampling (eqn.~\ref{eq:GPD-exponential-mu-sigma}) following HW2022. (d) MRL plot (section~\ref{sec:methods}\ref{sec:MRL}) for different precipitation thresholds. The blue shading defines a region of confidence calculated as $\pm$ one standard deviation. The red line denotes the threshold used in the GPD fit of HW2022 (i.e., the 20th percentile). The orange line represents the full linear equation from Equation~\ref{eq:MRL} ($\sigma+u\xi/(1-\xi)$).}
\label{fig:Madagascar-CFD-and-deviance}
\end{figure}

Fig.~\ref{fig:Madagascar-CFD-and-deviance}d shows the MRL plot (section~\ref{sec:methods}\ref{sec:MRL}) for the Madagascar Drought case. As a reminder, this plot needs to scale linearly within the range of the chosen threshold value (section~\ref{sec:methods}\ref{sec:MRL}) in order for the GPD analysis to be justified and self-consistent. For small and large values of precipitation, the MRL curve appears to curve nonlinearly, indicating that a GPD fit is likely invalid using those thresholds. For precipitation thresholds ranging from roughly 1.6 to 2 mm/day, the MRL curve plateaus. The orange line shows the expected MRL line, ($\sigma+u\xi/(1-\xi)$); that slope does not align with the MRL curve around the value of 1.9 corresponding to the 20th percentile precipitation threshold chosen by \cite{Harrington-Wolski-Pinto-et-al-2022:limited}.   In general, we suggest that this MRL analysis should be helpful in examining the self-consistency of applying a GPD to precipitation data, though the small number of data points appears an obstacle for a reliable MRL analysis in this particular case.  Specifically, it is not obvious if the required linear range exists to a degree that justifies using a GPD distribution and allows for selecting a threshold. 

We find again that an extreme value distribution is not required by the data distribution.  The GPD fit to the bottom 20th percentile precipitation data is no better than a log-normal distribution fitted to \textit{all} of the precipitation data (Fig.~\ref{fig:Madagascar-CFD-and-deviance}a).  Furthermore, both the log-normal and GPD fit indicate no need of a GMST dependence, in keeping with the small value of $\alpha$ found by HW2022. A log-normal distribution seems a reasonable null hypothesis for the positive precipitation data. In fact, the empirical CDF is practically unconstrained at its lower tail because there are no data points below a value of 1.25 mm/day. Supplementary Fig.~SI-6 for GPD, log-normal, and empirical distributions also suggests the log-normal distribution looks reasonable, albeit acknowledging the highly limited sample size in this case.

The deviance statistic for the Madagascar precipitation observations is marginally significant at the X\% level (leftmost green bar is above the red significance line 
, Fig.~\ref{fig:Madagascar-CFD-and-deviance}b), indicating that the inclusion of the additional model parameter $\alpha$ is justified.  Uncertainty plots for the GPD parameters fit to these precipitation observations are indicated in Supplementary Fig.~SI-7. The bootstrap estimates for the $\alpha$ parameter (Supplementary Fig.~SI-8a) ranges over both negative and positive values, indicating that it does not significantly differ from the null.  None of the 35 ensemble members exceed the deviance statistic significance level. Figure~\ref{fig:alpha_distributions_CTRL}c shows that there is no relation between these extreme drought events and GMST. One interpretation is that this is due to the absence of any strong effect of internal variability modes on both GMST and Madagascar drought, in contrast with the analysis of Australian temperatures in Figure~\ref{fig:alpha_distributions_CTRL}a. Also, Figure~\ref{fig:Madagascar-CFD-and-deviance}b may suggest that the CESM Large Ensemble does not adequately capture the effect of ACC on droughts in Madagascar.

As a reminder, the threshold parameter of the GPD, corresponding to the maximum precipitation value considered as an extreme drought year in this attribution case, was assumed by HW2022 to vary exponentially with the GMST ($\overline{T}$), as $u=u_0\exp(\alpha \overline{T}/u_0)$ with $u_0<0$. If droughts become more severe with increasing GMST, we expect the threshold parameter for the distribution of the negative precipitation to get larger (less negative) with increasing GMST, and therefore $\alpha/u_0$ to be negative and $\alpha$ to be positive. The value of ${\alpha}_\mathrm{GPD}$ calculated from observations is small and indistinguishable from zero (Fig.~\ref{fig:Madagascar-CFD-and-deviance}c). The exponential fit was assumed by HW2022 in order to ensure that the scaling factor $\exp(\alpha \overline{T}/u_0)$ applied to both $u_0$ and $\sigma_0$ is strictly positive, so the distribution has fixed dispersion \cite[][]{Philip-Kew-Van-et-al-2020:protocol}. A scatter plot of ERA5 precipitation against GMST anomalies looks notably linear (see the dashed blue line in Fig.~\ref{fig:Madagascar-CFD-and-deviance}c), consistent with the Taylor expansion of the exponential function, 
\begin{align*}
  u=u_0\exp(\alpha \overline{T}/u_0)
  \approx u_0(1+\alpha \overline{T}/u_0)
  =u_0+\alpha \overline{T}.
\end{align*}

The small value of $\alpha$ found for observations indicates that the increase in GMST has had little effect on the statistics of two-year-average precipitation in Madagascar. Whereas this is also the conclusion reached by HW2022, we found that the data can be fitted with a log-normal distribution rather than GPD and without GMST dependence (Fig.~\ref{fig:Madagascar-CFD-and-deviance}a). Further, the deviance statistic only marginally justifies adding GMST dependence for observations and does not justify adding such dependence for model output (Fig.~\ref{fig:Madagascar-CFD-and-deviance}b). The fact that only 14 data points are involved in the fit seems much too small to draw any meaningful conclusions on an extreme value distribution. Supplementary Fig.~SI-9a 
indicates that even if using the equivalent of 35 times as much data, the uncertainty range for $\alpha$ includes both positive and negative values. One could, therefore, conclude that there is no clear relationship between GMST and low rainfall in this region.

\section{Conclusions}
\label{sec:conclusions}

We were inspired in this work by a series of attribution studies by the ``World Weather Attribution initiative'' (WWA) and evaluated the part of their methodology that involves a fit of extreme value distribution functions to observations. The full protocol used by this effort \cite[][]{Philip-Kew-Van-et-al-2020:protocol} involves additional tools and model analyses, including selecting and defining extreme events, analyzing trends in observational data, using multi-model ensembles to simulate event scenarios with and without human-induced climate change, performing quality checks on model accuracy, and synthesizing results to determine climate change's influence on the event's intensity and likelihood. A central WWA method nevertheless involves fitting an extreme value distribution function to an observed record, with the distribution parameters depending on global mean surface temperature (GMST), whose variation is taken to represent ACC. For example, the location parameter of an extreme value distribution (GEV) is sometimes made a linear function of GMST, $\overline{T}$, as $\mu=\mu_0+\alpha\times\overline{T}$. The GMST dependence included in the function is then used to calculate the change in return time of extreme events. This method was applied to many extreme events such as heatwaves, cold spells, floods, droughts, and more \cite[e.g.,][]{Oldenborgh-Otto-Haustein-et-al-2016:heavy, Wiel-Kapnick-Oldenborgh-et-al-2017:rapid, Oldenborgh-Wiel-Sebastian-et-al-2017:attribution, Ciavarella-Cotterill-Stott-et-al-2021:prolonged, Oldenborgh-Krikken-Lewis-et-al-2021:attribution, Harrington-Wolski-Pinto-et-al-2022:limited}. We examined the robustness of the results of three example WWA studies, demonstrated when the results may be unreliable due to the effects of internal variability, and suggested several additional statistical tests, reviewed below, that may increase confidence in the results of such an analysis.

We began by examining the GMST dependence of extreme heatwave events in Siberia (following CC2021)  and Australia (following OK2021), as well as drought in Madagascar (following OK2021), using data from a CESM preindustrial simulation. We found a significant GMST dependence in the first two cases, indicating that such dependence can result from internal variability rather than only represent the effects of ACC. The GMST dependence, in these cases, is a result of regional internal variability such as ENSO or other climate variability modes, which affects the extreme events being examined, and also weakly affects GMST. This points that the value of $\alpha$ calculated in the more realistic case where both effects are present includes the effects of internal variability in addition to those of ACC. The value of $\alpha$ in the Australian case is especially large, and could naively be interpreted as indicating that every degree increase in GMST is associated with a 5-degree increase in Australian heatwaves. Heuristically, the large values of $\alpha$ can be thought of as a result of internal variability (possibly ENSO) affecting GMST by 0.2 degrees for every 1-degree effect it has on Australian heatwaves. As noted earlier, this issue could be mitigated in part by including indices for modes of internal variability as covariates in the model fitting.

We tested if an extreme value distribution is indeed needed by plotting the empirical cumulative distribution function (CDF) with bootstrapping error bars (Figs.~\ref{fig:Siberia_GEV_fit_and_deviance}a, \ref{fig:australia-CDF-and-deviance}a, \ref{fig:Madagascar-CFD-and-deviance}a). We found that fits using a standard distribution (normal or log-normal) fall within the error bars of the empirical distribution, indicating that the observations do not empirically allow for distinguishing between the characteristics of extreme value and normal distributions. The differing tails of these distributions, however, lead to different extreme event return times. Similarly, the standard distributions seem to fit the data well without a GMST dependence, indicating that such dependence may not be justified by the data. We further tested if the addition of a GMST dependence of the distribution parameters is justified statistically using the log-likelihood ratio test, also known as the deviance statistics. When adding a parameter --- in this case  $\alpha$ --- one expects the fit to improve, but the question is whether the improvement is beyond what is expected simply due to the addition of one more model parameter, so that one is convinced that the additional parameter represents a genuine effect in the data. We found that in the Australia and Madagascar cases, there is no justification for the addition of $\alpha$.

The Madagascar drought attribution (HW2022) was based on only fourteen of the lowest two-year-averaged precipitation rates, hardly sufficient to construct a meaningful empirical CDF or to constrain the four parameters of the GPD distribution used for that purpose. In addition, computing two-year averages of precipitation and applying an extreme value distribution may lead to inconsistencies a priori given that averages converge to a Gaussian distribution (under the Central Limit Theorem). The GPD fit to the record, in this case, is meant for high values above a threshold, and the expected mean needs to vary linearly with the threshold chosen. We follow \cite{Coles-Bawa-Trenner-et-al-2001:introduction} in suggesting that a mean residual life (MRL) plot can generally be used to test the self-consistency of using a GPD. In this case, it was not clear that these drought events are well represented by a GPD, but with the aforementioned overarching concern regarding the number of data points. To deal with the loss of data associated with this approach and difficulties in choice of an optimal precipitation threshold, \cite{Naveau-Huser-et-al-Precip} suggest instead using a statistical model fit to the full distribution of precipitation data, which has the added benefit of following the rules of extreme value theory for both high and low precipitation events. While our results---that attribution of two-year-averaged precipitation is not possible---agree with those of HW2022, we emphasized our different perspective: it is unlikely that attribution would be possible based on such a small number of data points, as well as the other difficulties pointed out above. In contrast, other approaches evaluating changes in the seasonal cycle of precipitation have led to the attribution of shifts in seasonal rainfall patterns in Southern Madagascar to climate change \cite[][]{Rigden-Golden-Chan-et-al-2024:climate}.

An important issue raised by this work is the discrepancy between the modeled and observed findings, as seen, for example, in the deviance statistic results for the observations versus CESM Large Ensemble (Figs.~\ref{fig:Siberia_GEV_fit_and_deviance}b, \ref{fig:australia-CDF-and-deviance}b, \ref{fig:Madagascar-CFD-and-deviance}b). This may be due to a particular sequence of internal variability in the observations, measurement error in the observational record, or issues with the simulation of extreme weather in the CESM Large Ensemble. It would be helpful to test if these discrepancies also occur with other models. The use of multiple models by WWA (\cite{Philip-Kew-Van-et-al-2020:protocol}, OK2021), may help mitigate against this kind of problem.

Overall, our results suggest uncertainty in the interpretation of the dependence of the parameters of extreme event distributions on the global mean surface temperature as reflecting ACC. The proposed and demonstrated deviance statistic, MRL plot, and a careful analysis of uncertainty ranges, especially for $\alpha$, may help avoid misguided confidence in attribution results. As noted in the introduction, WWA does not exclusively rely upon the empirical methodology evaluated here in order to reach conclusions, and also evaluates other historical data, model simulations, and the dynamical context. \cite{Philip-Kew-Van-et-al-2020:protocol} also highlight some caveats, including the necessity of testing for other probability laws (e.g., Gaussian, Gamma, etc.) and covariates (ENSO, PDO, etc.), among other issues. Nevertheless, the observations-based component of the WWA methodology examined here still figures prominently in their analyses, and we have highlighted several caveats with its interpretation, foremost that internal variability can give rise to the appearance of strong covariance between GMST and local extremes. It is important to study when the ACC signal emerges distinctly relative to internal variability on a case-by-case basis. The ACC signal is known to have emerged in most regions of the world for temperature extremes, while this may not be the case for variables like precipitation that have much higher variability. Time of emergence is a factor that should always be considered when discussing the results of an attribution study. Our results do not provide evidence against a relationship between GMST and local extremes but, rather, that internal variability needs to be taken into account when fitting a distribution function to the record. As currently implemented, the WWA approach can lead to first-order biases in some cases. It, therefore, seems that it may be best to use a multi-method approach to ensure a robust attribution approach \cite[e.g.,][]{shepherd2018storylines, leach2024heatwave, robin2020nonstationary, sippel2024could}.

We suggest two key takeaways. First, controlling for internal variability, possibly by making the distribution parameters depend on additional covariates such as ENSO, appears important in order to accurately estimate the mean ACC trend. This improvement upon the current method should be further developed. Inclusion of additional covariates is vital as a way to avoid biases in the estimate of ACC effects on extreme event statistics. Second, empirical attribution techniques should be examined by applying the approach to simulations with and without changes to anthropogenic emissions. Application to such simulations would make it much easier to identify the likely influence of other factors on the local response in the absence of ACC, and so to ensure that all relevant modes are included in the statistical model, assuming the models are right in this respect.

\acknowledgments
PS is supported by the Harvard Global Institute. ET is supported by DOE grant DE-SC0023134, and thanks the Weizmann Institute for its hospitality during parts of this work. PH is supported by NSF Award 2123295. The CESM project is supported primarily by the National Science Foundation (NSF). We acknowledge the CESM Large Ensemble Community Project and supercomputing resources provided by NSF/CISL/Yellowstone. We also acknowledge instructive conversations with Angela Rigden.

\datastatement
All data used in this paper are available in public repositories. The climatological output from the CESM Large Ensemble and CESM preindustrial runs used in this paper are publicly available using the \href{https://intake-esm.readthedocs.io/en/stable/}{Intake-ESM} Python package. All codes and data used here are available under the public open science foundation depository \href{https://osf.io/fycu2/}{https://osf.io/fycu2/}, DOI 10.17605/OSF.IO/FYCU2.

\newpage
\bibliographystyle{ametsocV6}
\bibliography{export}

\clearpage
\newpage

\setcounter{figure}{0}    
\setcounter{section}{0}    
\renewcommand\thefigure{SI-\arabic{figure}}    
\renewcommand\thesection{\Alph{section}}    
\end{document}